%% file: main.tex
\documentclass[letterpaper, 10 pt, conference]{ieeeconf}  

\IEEEoverridecommandlockouts                      

\usepackage{graphics} 
\usepackage{caption}
\usepackage{subcaption}
\usepackage{epsfig} 
\usepackage{amsmath} 
\usepackage{amssymb}  
\usepackage{amsfonts}
\usepackage{physics}
\usepackage{cite}
\usepackage[hidelinks]{hyperref}
\hypersetup{
	colorlinks=false,
	linkcolor=blue,
	filecolor=cyan,      
	urlcolor=magenta,
	citecolor=blue,
	pdfpagemode=FullScreen,
}
\usepackage{algorithmicx, algorithm, algpseudocode}
\usepackage{xcolor}
\usepackage{booktabs}

\makeatletter
\let\NAT@parse\undefined
\makeatother

\usepackage[capitalise,nameinlink,sort]{cleveref}

\makeatletter
\newcounter{algorithmicH}% New algorithmic-like hyperref counter
\let\oldalgorithmic\algorithmic
\renewcommand{\algorithmic}{%
	\stepcounter{algorithmicH}% Step counter
	\oldalgorithmic}% Do what was always done with algorithmic environment
\renewcommand{\theHALG@line}{ALG@line.\thealgorithmicH.\arabic{ALG@line}}
\makeatother

\let\labelindent\relax
\usepackage[inline]{enumitem}

\newtheorem{theorem}{\bf Theorem}

\newtheorem{lemma}{\bf Lemma}
\newtheorem{assumption}{\bf Assumption}
\newtheorem{definition}{\bf Definition}

\title{\LARGE \bf
Online Residual Learning from Offline Experts for Pedestrian Tracking
}

\author{Anastasios Vlachos, Anastasios Tsiamis, Aren Karapetyan, Efe C. Balta, and John Lygeros% <-this % stops a space
\thanks{This work has been supported by the Swiss National Science Foundation under NCCR Automation (grant agreement 51NF40 180545), and by the  European Research Council under the ERC Advanced grant agreement  787845 (OCAL).}
\thanks{A. Vlachos, A. Tsiamis, A. Karapetyan, and J. Lygeros are with the Automatic Control Laboratory, Department of Information Technology and Electrical Engineering, ETH Z{\"u}rich, 8092 Z{\"u}rich, Switzerland (e-mail: avlachos@student.ethz.ch, \{atsiamis, akarapetyan, jlygeros\}@control.ee.ethz.ch). E. C. Balta is with the Control and Automation Group, inspire AG, 8005 Zürich, Switzerland, and with the Automatic Control Laboratory {(e-mail: efe.balta@inspire.ch)}.}%
}

\begin{document}

\maketitle
\thispagestyle{empty}
\pagestyle{empty}

%%%%%%%%%%%%%%%%%%%%%%%%%%%%%%%%%%%%%%%%%%%%%%%%%%%%%%%%%%%%%%%%%%%%%%%%%%%%%%%%
\begin{abstract}
In this paper, we consider the problem of predicting unknown targets from data. 
We propose Online Residual Learning (ORL), a method that combines online adaptation with offline-trained predictions. At a lower level, we employ multiple offline predictions generated before or at the beginning of the prediction horizon. We augment every offline prediction by learning their respective residual error concerning the true target state online, using the recursive least squares algorithm. At a higher level, we treat the augmented lower-level predictors as experts, adopting the Prediction with Expert Advice framework. We utilize an adaptive softmax weighting scheme to form an aggregate prediction
and provide guarantees for ORL in terms of regret. We employ ORL to boost performance in the setting of online pedestrian trajectory prediction. Based on data from the Stanford Drone Dataset, we show that ORL can demonstrate best-of-both-worlds performance.
\end{abstract}

\input{sections/introduction}
\input{sections/formulation}

\input{sections/algorithm}

\input{sections/analysis}

\input{sections/simulations}

%%%%%%%%%%%%%%%%%%%%%%%%%%%%%%%%%%%%%%%%%%%%%%%%%%%%%%%%%%%%%%%%%%%%%%%%%%%%%%%%
\section{Conclusion and Future Work}
We propose Online Residual Learning as a way of bridging online adaptation with offline-trained models. It applies online regression on the residual errors of the offline predictions and then it combines them online in a Multiple Experts Learning setting. We provide regret guarantees to show that our method is competitive against the best in hindsight residual error predictor with bounded path length. Future work will study formal strategies for the choice of residuals  
as well as the development of feedback controllers based on our framework. Note that the aggregate prediction~\eqref{eq:aggregate_prediction} is naive, in the sense that it does not incorporate any prior information, for example, the position of obstacles. Exploiting structure and side information could potentially further improve online prediction performance.

%%%%%%%%%%%%%%%%%%%%%%%%%%%%%%%%%%%%%%%%%%%%%%%%%%%%%%%%%%%%%%%%%%%%%%%%%%%%%%%%

\bibliographystyle{IEEEtran}
\bibliography{bibliography}
\appendix
\input{sections/appendix}
\end{document}

%% file: sections/introduction.tex
\section{Introduction}\label{sec:introduction}
Making predictions from data is a fundamental process in many fields, such as finance \cite{lin2011machine}, robotics \cite{demiris2007prediction}, autonomous driving \cite{cui2019multimodal}, and pedestrian safety \cite{ridel2018literature}. In many modern applications, the assumption that the model generating the data we wish to predict is known may not hold; the model may be unknown or changing with time. Furthermore, data may become available in a streaming fashion due to constraints. The learning architectures that tackle this problem fit into one of two paradigms: offline learning and online learning.

Offline Learning comprises \emph{static} methods that aim to build a model for the target we wish to predict, using historical data. In many applications of interest, offline data may be widely available. Typically, when learning is performed offline, we can also afford more time for training complex models. Predictions generated by offline models can be highly accurate, especially in cases where the data we wish to predict comes from the same distribution as the data used during training. However, offline predictions might perform poorly under distribution shifts. 
On the other hand, online learning involves \emph{dynamic} methods that rely mainly on data gathered on the go. Online learning can adapt to abrupt changes or distribution shifts. However, it might exhibit poor transient performance, due to a lack of prior knowledge. 

In this work, we propose Online Residual Learning (ORL), a method that exploits the offline predictions in the online learning setting. Our contributions are the following.

\textbf{Online learning over offline predictions}. We incorporate offline predictions to improve online performance. At a lower level, we learn online the residual errors of the offline predictions, using the recursive least squares algorithm. At a higher level, we aggregate the lower level predictions treating them as \emph{experts} during online estimation, a standard technique in online learning \cite{cesa2006prediction, yuan2020trading}. Our approach acts as a ``wrapper'' around offline models, without modifying them. Hence, it avoids the problem of catastrophic forgetting, a common issue in domain adaptation \cite{song2022hybrid}. 

\textbf{Regret Guarantees.} We provide performance guarantees in terms of regret. We show that ORL competes against the best, in hindsight, offline prediction, equipped with the best, in hindsight, residual predictor in a restricted class. This class includes arbitrary linear time-varying predictors with bounded path length, that is, bounded total variation.

\textbf{Pedestrian prediction.} We apply ORL to the prediction of pedestrian trajectories. We use the data from the Stanford Drone Dataset while the offline predictions are generated by the offline trained model of~\cite{Mangalam_2021_ICCV}. We observe that ORL can surpass the performance of both purely offline and purely online methods. The offline predictions improve the transient performance, while we still adapt to the individual pedestrian's motion.

Our approach is inspired by boosting methods, which have been popular both in the offline~\cite{freund1996experiments} (i.e. AdaBoost) and in the online setting~\cite{beygelzimer2015online} as well as in control applications~\cite{agarwal2020boosting}, and learning with multiple experts, which has ubiquitous applications~\cite{seldin2013evaluation, gadginmath2023fusing, krajewski2024scaling}. Recently, attempts have been made to combine offline and online learning approaches, especially in the context of Reinforcement Learning \cite{song2022hybrid, wagenmaker2023leveraging}. 

To learn the residual errors online, we appeal to online convex optimization techniques~\cite{hazan2007logarithmic}. In particular, we employ the widely used recursive least squares algorithm with forgetting factors~\cite{yuan2020trading}. By tuning the forgetting factor we can trade between static and dynamic regret~\cite{yuan2020trading}. Residual learning has also been employed for closed-loop control. In this setting, a learning model is trained offline or tuned to predict the mismatch between the nominal control model and the true process outputs. Mismatch learning-based control has been demonstrated, for example, on safe autonomous driving with model predictive control \cite{hewing2019cautious,mckinnon2019learn,kabzan2019learning} and precision motion control with iterative learning control~\cite{balta2021learning}. 

Pedestrian trajectory prediction, together with its counterpart problem of pedestrian intention prediction, has been well-studied in the literature. Works have been conducted on predicting intention \cite{chen2021psi}, predicting trajectories \cite{Mangalam_2021_ICCV, yuan2021agentformer}, or combining both approaches to boost performance \cite{rasouli2019pie}. We refer the interested reader to \cite{zhang2023pedestrian} for an extensive review. 

In \cite{yao2023eanet}, the authors propose to use an attention mechanism to learn online from offline predictions of pedestrian trajectories. Adapting complex models online can be challenging due to increased computational loads, contrasting our approach where the online adaptation occurs over simple models. 

Human motion prediction, in general, has become increasingly important in environments where humans interact with intelligent autonomous systems, and therefore, this problem has attracted attention from the research community. In \cite{butepage2018anticipating} a variational autoencoder, which is an offline model, is proposed, that can be used for online motion prediction. Another approach is that of \cite{bajcsy2020robust, bajcsy2021analyzing}, where authors aim to learn online and robustly predictive human models by employing tools from reachability analysis and optimal control theory. Our framework, differs from the Human-Robot Interaction setup, since we do not consider the problem of an autonomous vehicle interacting with a pedestrian in an urban environment. 

\textbf{Notation. }Let $\norm{x}$ denote the Euclidean norm for vectors $x\in\mathbb{R}^n$. Let $\norm{M}$ denote the spectral norm of a matrix $M$, and $\norm{M}_F$ the Frobenius norm. For positive definite $P \succ 0$, the weighted Frobenius norm is defined as $\norm{M}_{F,P} = \sqrt{\Tr(MPM^{\top})}$, where $\Tr{\cdot}$ denotes the trace of a matrix. For a given $n\in\mathbb{N}^{+}$, we define the identity matrix of dimension $n$ as $I_n$.

%% file: sections/formulation.tex
\section{Problem Setting}\label{sec:formulation}
Consider an \emph{unknown} target, e.g. a pedestrian, moving according to unknown autoregressive dynamics
\begin{equation}
    r_{t+1} = f(r_{t-p+1:t})+d_t,
    \label{eq:data_gen}
\end{equation}

where $r_t\in\mathbb{R}^n$ is the target state, $d_t$ is a bounded disturbance perturbing the state, $f:\mathbb{R}^{np}\to\mathbb{R}^n$ is the dynamics, $p$ is the memory of the dynamics, and $r_{t-p+1:t}$ contains the past $p$ target states: 
\begin{equation}
    r_{t-p+1:t} = \begin{bmatrix}
        r_{t-p+1}^{\top} & r_{t-p+2}^{\top} & \dots & r_t^{\top}\ 
        % r_{t-p+2} \\
        % \vdots \\
        % r_{t}
    \end{bmatrix}^{\top}.
\end{equation}

We are interested in predicting the target state $k$-steps into the future, for some $k\ge 1$. Since the dynamics are unknown, we need to use data to learn how to predict and adapt to the target's trajectory online.

For simplicity, we present the case $k=1$ in the main text and provide the general case of $k>1$ in~\cref{app:k-steps ahead RLS}. 

Assume we deploy a prediction algorithm at time $t=0$ up to some terminal time $T$. Let $\hat{r}_{t+1}\in\mathbb{R}^n$ denote the one-step-ahead prediction of the target at time $t$. In our setting, the following occur sequentially at each time step $t\geq 0$:
\begin{enumerate}
    \item The current target state $r_t$ is received and a prediction error $\ell_t(\hat{r}_t)\triangleq\|r_t-\hat{r}_t\|^2$ is incurred.
    \item The predictor is updated based on the current and past target states.
    \item A prediction 
    $\hat{r}_{t+1}$ is made.
\end{enumerate}

While this formulation fits the framework of online regression, we depart from the purely online setting, where we rely exclusively on the sequentially gathered data. We assume to have additional access to $N$ \emph{offline} trajectory predictions $\hat{r}_{0:T,i}^{\text{off}} = \{\hat{r}_{t,i}^{\text{off}}\}_{t=0,1,\dots,T}$, $i\in \{1,2,\dots,N\}$. These are given before deployment of the prediction algorithm and are generated based on the initial conditions $r_{-1},\dots,r_{-p}$\footnote{Our algorithm can also accommodate recursively generated predictions of the form $\hat r^{\mathrm{off}}_{t+1}=\hat{f}(r_{t-p+1:t})$. In the pedestrian tracking problem, we face the former case.}. We also use the term \emph{expert} to refer to the offline predictions~\cite{cesa2006prediction}. The offline predictions can be generated from models that have been trained offline based on historical data. In many applications of interest, including pedestrian tracking, such historical data are widely available. We consider the offline predictions as given and do not investigate how the models are obtained or trained. 

Our goal is to combine the offline predictions with the sequentially gathered online data to boost the overall online prediction performance. Specifically, we aim to use the prior knowledge of offline predictions as a warm start for our online learners to improve prediction performance.

Ideally, the online predictor should minimize the cumulative prediction error
\begin{equation}
    \mathcal{L}_T \triangleq \sum_{t=1}^T \ell_t(\hat{r}_t).
\end{equation}
However, finding the optimal prediction policy is difficult, especially for arbitrary disturbances $d_t$ and unstructured dynamics~\eqref{eq:data_gen}. Instead, we consider the notion of \emph{Regret}, where we compare our performance against a restricted but meaningful class of benchmark prediction policies. 

Let $z_{1},\dots,z_T$ be a sequence of predictions generated by some benchmark policy class to be determined later. The regret is defined as 
\begin{equation}
    \mathcal{R}_T = \sum_{t=1}^T \left(\ell_t(\hat{r}_t) -  \ell_t(z_t)\right).
\end{equation}
Typically, the benchmark policy is allowed to be optimal in hindsight, i.e., it has access to all target states in advance. 

To guarantee the online prediction problem is well-behaved, we make the following boundedness assumption.

\begin{assumption}[Boundedness]\label{assum:bounded_residuals}
 For every expert $i$, $i=1,\dots,N$ the respective offline prediction error is uniformly bounded. That is, there exists a constant $D_r>0$ such that for any terminal time $T$
    \[
    \norm{r_t-\hat{r}^\mathrm{off}_{t,i}}\le D_r,\,\text{for all } 0\le t\le T.
    \]
\end{assumption}

~\cref{assum:bounded_residuals} is standard in online learning frameworks and it can be satisfied if the ground-truth trajectories and the offline predictions themselves are bounded. In the example of pedestrian trajectory prediction, both are satisfied since we are searching over a compact set (pixel space). Note that no other assumption is required for system~\eqref{eq:data_gen}, which is allowed to be arbitrary and that we should know at least an upper bound for $D_r$.

%% file: sections/algorithm.tex
\section{Online Residual Learning}\label{sec:algorithm}
\begin{figure}[H]
    \centering
    \includegraphics[width=\columnwidth]{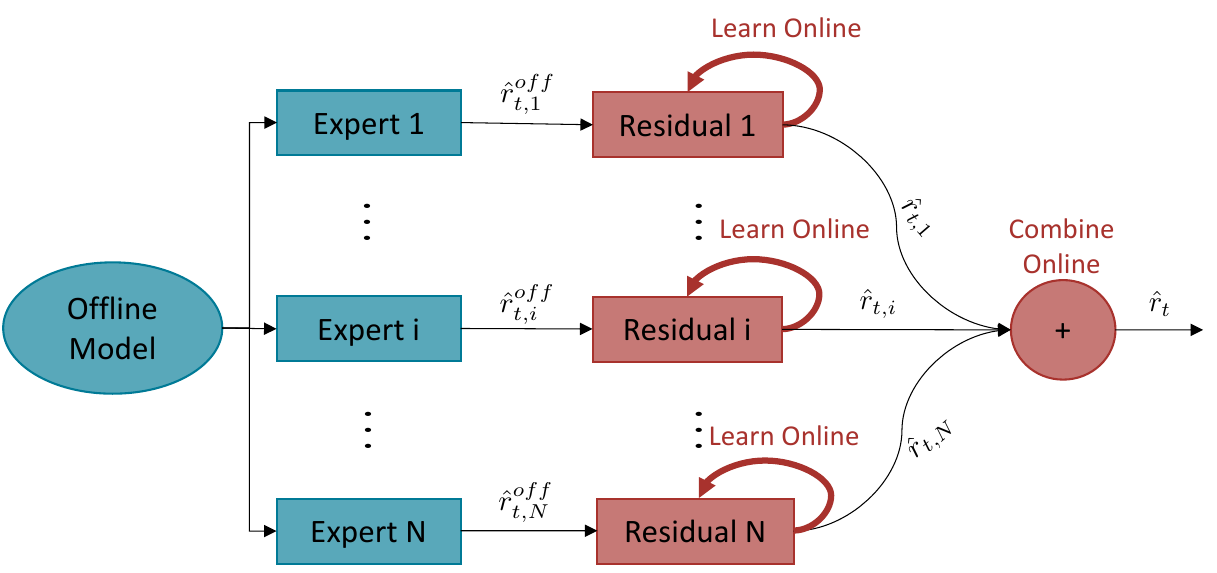}
        \caption{Schematic of the Online Residual Learning method. We attempt to learn the residual errors of the offline experts and combine them online.}
    \label{fig:res-off-on}
\end{figure}

Our proposed method, Online Residual Learning from Offline Experts, or simply Online Residual Learning (ORL), combines two common approaches in online learning: online regression and learning with expert advice. ORL is decomposed into two levels as shown in \cref{fig:res-off-on}. At the lower level, we employ the recursive least squares algorithm to predict the residual errors between the true target states and the offline predicted ones. We then use the predicted residual errors to correct the offline predictions. At the higher level, we treat the lower-level predictors as experts. We employ a standard softmax meta-algorithm to form an aggregate prediction that adapts to the best expert.  
\subsection{Residual Error Learning}

Define the residual errors between the offline predictions and the true target state as:
\begin{equation}
    e_{t,i} \triangleq r_t - \hat{r}_{t,i}^{\mathrm{off}} 
    \label{eq:res_error}
\end{equation}
where $r_t$ is the (sequentially revealed) target state and $i\in\{1,2,\dots,N\}$ is the expert index. The main idea is to predict the residual errors $e_{t,i}$. By considering the residual errors, rather than the raw target state $r_t$ directly, we can incorporate the prior knowledge coming from the offline predictions. Meanwhile, the offline predictions on their own provide good initial guesses, but they do not adapt to the target's motion. 

Let $\hat{e}_{t+1,i}$ denote the prediction of the one-step ahead residual error at time $t$ for the $i$-th expert. Then, at every time $t$, every expert $i$ outputs the corrected prediction 
\begin{equation}\label{eq:augmented_expert_prediction}
\hat{r}_{t+1,i}\triangleq\hat{e}_{t+1,i}+\hat{r}^\mathrm{off}_{t+1,i},\,i\in\{1,2,\dots,N\}. 
\end{equation}
The loss for each expert can then be written as
\[
\ell_t(\hat{r}_{t,i})=\|r_t-(\hat{e}_{t,i}+\hat{r}^{\mathrm{off}}_{t,i}) \|^2 =\|  e_{t,i}-\hat{e}_{t,i}\|^2.
\]
Hence, predicting the target $r_t$ is equivalent to predicting the residual errors. 

ORL predicts the residual errors using past information. It employs the Recursive Least Squares (RLS) algorithm with forgetting factors, which aims to find the best possible linear predictor, while being able to adapt to changing dynamics. Define the vector of past regressors at time $p$ as
\begin{equation}\label{eq:past_regressors}
       z_{t,i}= \begin{bmatrix}
        e_{t-p+1,i}^{\top} & e_{t-p+2,i}^{\top} & \dots & e_t^{\top}\ 
    \end{bmatrix}^{\top}.
\end{equation}
At every time, every expert $i\in\{1,2,\dots,N\}$ minimizes the discounted cumulative prediction error along with a regularization penalty term
\begin{equation}
\begin{aligned}
 \hat{M}_{t+1,i}&=\mathop{\mathrm{arg\,min}}_{M\in\mathcal{S}}   J_t(M)+\gamma_i^t\varepsilon_{i} \norm{M}_F^2,\\
 J_t(M)&\triangleq \sum_{\tau=1}^t \gamma_i^{t-\tau} \norm{e_{\tau,i} - M z_{\tau-1,i}}^2,
    \label{eq:cost}
    \end{aligned}
\end{equation}
where $\varepsilon_{i}>0$ is the regularization multiplier, and $\gamma_i\in(0,1]$ is the forgetting factor. Finally, the expert predicts
\begin{equation}\label{eq:RLS_prediciton}
\hat{e}_{t+1,i}=\hat{M}_{t+1,i}z_{t,i}.
\end{equation}

Note that we constrain the decision variable $M$ to lie in the set $\mathcal{S} \triangleq \{M\in\mathbb{R}^{n\times np}:\: \norm{M}\leq D\}$, for some hyperparameter $D\geq 0$. The constraint provides explicit control on the norm of $M$, guaranteeing bounded predictions $\hat{e}_{t,i}$.
A recursive update with projections is given in \cref{alg:RLS}. While we assume the same memory $p$ for the past regressors as in the target dynamics in~\eqref{eq:data_gen}, we may use a different one as well. In practice, we treat $p$ as a hyperparameter.

We note that the RLS algorithm does not require the underlying process, in our case $e_{t,i}$, to obey linear dynamics.  As discussed in \Cref{sec:regret_guarantees}, RLS achieves a competing performance (in terms of regret) when compared against other \emph{linear} predictors -- even if the residual errors evolve in a non-linear fashion. The only requirement for the RLS algorithm to be well-behaved is \cref{assum:bounded_residuals}.

\begin{algorithm}
    \caption{One-step ahead RLS prediction}
    \label{alg:RLS}
    \begin{algorithmic}[1]
        \Require Forgetting factor $\gamma_i \in (0,1]$ for expert $i\in\{1,2,\dots,N\}$, regularizer $\varepsilon_i$
        \State Initialize $\hat{M}_{0, i}\in S$ 
        \State Initialize $P_{0,i} =\varepsilon_i I_{np}$
        \For{$t=1,\dots,T$} 
        \State Make prediction $\hat{e}_{t,i} = \hat{M}_{t,i}z_{t-1,i}$
        \State Receive true state $r_t$, compute residual error $e_{t,i}$ 
        \State Incur loss $\ell_t(\hat{M}_{t,i}) = \norm{e_{t,i} - \hat{e}_{t,i}}^2$
        \State Update $P_{t,i} = \gamma_i P_{t-1, i} + z_{t-1,i} z_{t-1,i}^{\top}$
        \State $\hat{M}_{t+1, i}^{\star} = \hat{M}_{t, i} + (e_{t,i} - \hat{M}_{t, i} z_{t-1, i})z_{t-1, i}^{\top} P_{t, i} ^{-1}$
        \State Project $\hat{M}_{t+1, i} = \mathop{\mathrm{arg\,min}}_{M\in S}\norm{M-\hat{M}_{t+1, i}^{\star}}_{F, P_{t,i}}$
        \EndFor
        
    \end{algorithmic}
\end{algorithm}

\subsection{Combining Multiple Predictions}
Since we obtain one prediction per expert, we need a meta-algorithm for aggregating the predictions into a single one. This is exactly the goal of learning with expert advice~\cite{cesa2006prediction}. Specifically, at any time $t$, we assign weights $w_{t,i}$, representing the levels of trust, to each expert $i$, $i=1,\dots,N$. These weights are updated online using a softmax function, based on the prediction errors (losses) of each expert
\begin{equation}\label{eq:MOE_weight_update}
       w_{t+1,i} = \frac{w_{t,i} \exp(-\lambda \ell_{t}(\hat{r}_{t,i}))}{\sum_i w_{t,i} \exp(-\lambda \ell_{t}(\hat{r}_{t,i}))},
\end{equation}
where we initialize all weights to $1/N$.
Finally, the aggregate prediction is the weighted average of all experts' predictions
\begin{equation}\label{eq:aggregate_prediction}
    \hat{r}_t = \sum^N_{i=1} w_{t,i} \hat{r}_{t,i}.
\end{equation}
The whole procedure is summarized in \cref{alg:meta_alg}. 
Parameter $\lambda$ is a learning rate that determines how fast the weights can change. As shown later in Theorem~\ref{thm:regret_guarantees}, if $\lambda$ is chosen to be small enough, we can track the best experts and avoid overfitting to outliers. 
In particular, the larger the magnitude $D_r$ of the residual errors is, the smaller $\lambda$ should be to reduce sensitivity to outliers. 

\begin{algorithm}
    \caption{Prediction with Expert Advice}
    \label{alg:meta_alg}
    % \footnotesize
    \begin{algorithmic}[1]
        \Require Learning rate $\lambda$, set of $N$ experts $\mathcal{E}_N = \{E_i|i=1,\dots,N\}$
        \State $w_{1,i}= \frac{1}{N}$ for every Expert $i\in\{1,2,\dots,N\}$.
        \For{$t=1,2,\dots,T$} 
        \State Obtain prediction $\hat{r}_{t,i}\! =\! \hat{r}_{t,i}^{\mathrm{off}} + \hat{e}_{t,i}$ from each Expert~$i$.
        % \State $\hat{r}_{t,i} = \hat{r}_{t,i}^{\mathrm{off}} + \hat{e}_{t,i}$ for each Expert $i$.
        \State Aggregate prediction $\hat{r}_t = \sum_i w_{t,i} \hat{r}_{t,i}$
        \State Incur loss $\ell_{t}(\hat{r}_{t,i})$ for each Expert $i$
        \State Update the weights as 
        \begin{equation*}
            w_{t+1,i} = \frac{w_{t,i} \exp(-\lambda \ell_{t}(\hat{r}_{t,i}))}{\sum_i w_{t,i} \exp(-\lambda \ell_{t}(\hat{r}_{t,i}))}
        \end{equation*}
        \EndFor
    \end{algorithmic}
\end{algorithm}

%% file: sections/analysis.tex
\section{Regret Guarantees}\label{sec:regret_guarantees}
In this section, we present theoretical guarantees for ORL in terms of regret bounds, where we compare ORL with a set of benchmark policies. 

Let us define the benchmark policy class more formally. Fix an expert $i$, $i=1,\dots,N$ and let $M_{t,i}\in\mathbb{R}^{n\times np}$, $t=1,\dots,T$ be any sequence of predictor matrices, also called a comparator sequence. Note that we assume comparators with the same memory $p$ as in~\eqref{eq:cost}.
Let 
\[
\tilde{r}_{t+1,i}(M_{t+1,i})\triangleq M_{t+1,i}z_{t,i}+\hat{r}^{\mathrm{off}}_{t+1,i}
\]be the predicted target state if $M_{t+1,i}$ is used in place of $\hat{M}_{t+1,i}$. Abusing the notation, let 
\[
\mathcal{L}_T(M_{1:T,i})\triangleq \sum_{t=1}^{T}\ell_t(\tilde{r}_{t,i}(M_{t,i}))
\]
denote the respective cumulative error. 
We consider sequences of bounded path length
\begin{equation}\label{eq:path_length}
    V_{T,i}\triangleq \sum_{t=1}^{T-1} \norm{M_{t+1,i} - M_{t,i}}_F.
\end{equation}
Given an upper bound $V_T$ on the path length, the regret is defined as
\begin{equation}\label{eq:formal_regret}
    \mathcal{R}_{T}(V_T)=\mathcal{L}_T-\min_{i=1,\dots,N}\mathop{\min_{M_{t,i}\in\mathcal{S}}}_{V_{T,i}\le V_T}\mathcal{L}_T(M_{1:T,i}).
\end{equation}

Since the benchmark prediction policy minimizes the cumulative error over all time steps, it is optimal in hindsight. As we allow larger path lengths, the regret should increase since the comparator sequences are allowed to be more expressive. On the other hand,
if we choose $V_T=0$, then we compare against the optimal in hindsight, \emph{static} time-invariant predictors $M_{t+1,i}=M_{t,i}$.
We have the following regret guarantees.
\begin{theorem}[Regret guarantees]\label{thm:regret_guarantees}
    Consider the ORL prediction, which is given by~\eqref{eq:augmented_expert_prediction},~\eqref{eq:RLS_prediciton}, and~\eqref{eq:aggregate_prediction}. Choose the learning rate $\lambda$ of~\cref{alg:meta_alg} such that 
    \begin{equation}\label{eq:lambda_bound}\lambda \leq \frac{1}{4(D_r^2 + D^2D_r^2) }.
    \end{equation} Let $\varepsilon_i=1, i=1,\dots,N$
    and fix a path-length budget $V_T$. Choose forgetting factors for each expert $i\in\{1,2,\dots,N\}$ equal to \begin{equation}\label{eq:forgetting_factor}\gamma_i = 1 - \frac{1}{2}\sqrt{\frac{\max \{V_{T}, \log^2T/T\}}{2DT}},
    \end{equation}
    where $D$ is the norm bound on $M$ in set $\mathcal{S}$.
    Then, 
    \begin{align*}
  \mathcal{R}_T(V_T)\le  \max\{\mathcal{O}(\log T), \mathcal{O}(\sqrt{TV_{T}})\} + \mathcal{O}(\log N),
    \end{align*}
  where $\mathcal{O}(\cdot)$ keeps track of terms involving $T$ and $N$.
\end{theorem}
The proof, given in~\cref{proof:thm:regret_guarantees}, is based on the results of~\cite{yuan2020trading}. As we compete with benchmarks that vary more with time (higher path-length $V_T$), the forgetting factor should decrease so that we forget faster. Intuitively, this is because we need to adapt faster to changes. Conversely, to compete with static benchmarks, we should choose a forgetting factor close to $1$. The bound also reflects the intuition that a higher path length leads to higher regret. 

As a sanity check, we investigate the special case when the path length is zero, where we compete against static predictors. This class contains some important naive baselines, for example, $M_{t,i}=0$ or $M_{t,i}=\begin{bmatrix}
    I_n&0&\cdots&0
\end{bmatrix}$.
The former captures the case when we predict $\hat{r}_{t,i}=\hat{r}^{\mathrm{off}}_{t,i}$, that is, when we disregard the residual errors and fully trust the offline predictions. The latter one captures the case when we predict $\hat{e}_{t+1,i}=e_{t,i}$, i.e. when we believe that the residual error is constant. In both cases, if we tune the forgetting factors according to~\eqref{eq:forgetting_factor}, then we get a regret guarantee of
 \begin{align*}
  \mathcal{R}_T(V_T)\le  \mathcal{O}(\log T) + \mathcal{O}(\log N).
    \end{align*}
Since we have only logarithmic terms, the regret is negligible in this case. Hence, with proper tuning, ORL is guaranteed to perform at least as well as the best naive baselines.

%% file: sections/simulations.tex
\section{Pedestrian Tracking} \label{sec:pedestrian_tracking}
In this section, we present simulation results that showcase how ORL performs on the problem of Pedestrian Trajectory Prediction. In the beginning $(t=0)$, we acquire the offline predictions from the offline-trained model of \cite{Mangalam_2021_ICCV} and then we deploy our proposed learning algorithm. We opt to not retrain the offline models on the go, due to computational complexity. 

\subsection{Dataset} 
We choose to experiment on the \textit{Stanford Drone Dataset} (SDD) \cite{robicquet2016learning}. It contains more than 40,000 trajectories of many street entities, like cars, buses, and bicycles, a quarter of which correspond to pedestrians. These trajectories are captured from a top-down view of 20 different scenes across the Stanford campus. The dataset is sampled at 30 FPS.

As pre-processing, we filtered out short trajectories and the ones that do not correspond to pedestrians, as in \cite{Mangalam_2021_ICCV}. We did not undersample the trajectories, compared to [8], since we observed that this degrades the performance of our method. Furthermore, we use the center of the bounding boxes, to get a single pixel coordinate representation for each pedestrian.

\subsection{Offline Model}
We adopt the model proposed in \cite{Mangalam_2021_ICCV} as the offline model for our simulations. The main idea of this model is to split the overall multimodality of the prediction problem into two factors: the uncertainty on the final goal (\emph{epistemic} uncertainty) and the uncertainty on the path followed towards the final goal, for a known final goal (\emph{aleatoric} uncertainty). The model takes as input initial trajectory data $r_{-1}, \dots r_{-K_e}$, where $K_e$ is a hyperparameter of the offline model. It predicts future motion by producing multiple predictions offline; it samples multiple points from the scene as final trajectory points and, for each of them, it predicts the pedestrian trajectory from the starting point to this final point.

\subsection{Simulation Results}
We run the pre-trained model of \cite{Mangalam_2021_ICCV} on a Linux operating system, with Python 3.8.3 and PyTorch with Cuda 12.1 on a GeForce RTX 3080 GPU.
For simulations, we choose the forgetting factors $\gamma_i = 0.8$ and the memory $p=2$ for all experts $i\in\{1,2,\dots,N\}$ and the learning rate of~\cref{alg:meta_alg} $\lambda = 10^{-4}$.

We present results for two individual pedestrians on two different scenes, namely the `Hyang0' and the `Hyang3', as they are termed in the SDD. More specifically, we present instantaneous and cumulative error plots, as well as predicted trajectory plots on each scene. In these plots, the prediction resulting from our proposed method is compared to the following methods listed in~\cref{tab:pred_methods}: i) the \emph{online} method, where we perform RLS prediction directly on the target $r_t$ without considering any offline experts, ii) the \emph{offline experts} method, where we consider directly the raw offline predictions as experts, and we combine them online using~\cref{alg:meta_alg}, as well as iii) the \emph{best offline} one which returns the prediction with the smallest \emph{Average Displacement Error} (ADE): $\mathrm{ADE}_i =  \frac{1}{T} \sum_{t=1}^T \ell_t(\hat{r}_{t,i})$.

\begin{figure}[H]
        \centering
        \begin{subfigure}{0.85\columnwidth}
            \centering
            \includegraphics[width=\columnwidth]{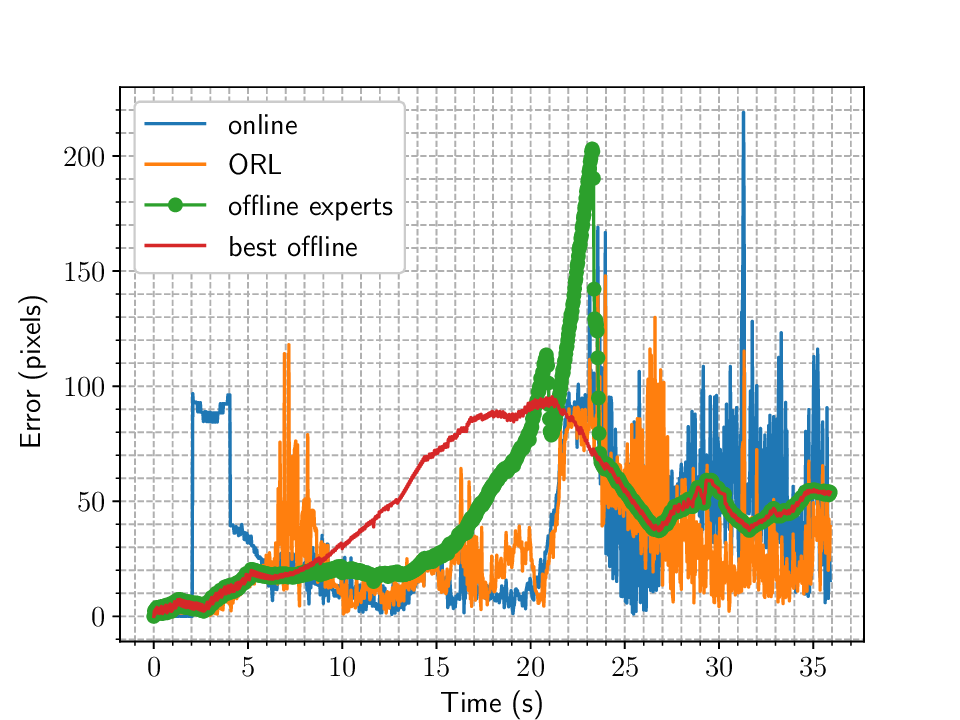}
            \caption{Instantaneous Error Plot}   
            \label{}
        \end{subfigure} 
        
        \begin{subfigure}{0.85\columnwidth}  
            \centering 
            \includegraphics[width=\columnwidth]{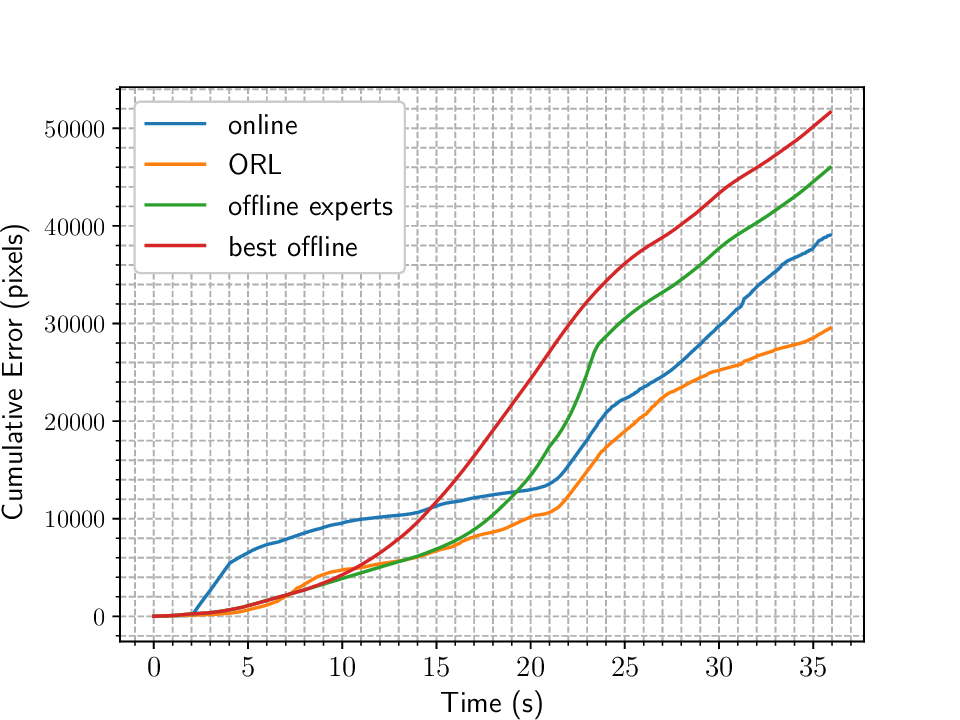}
            \caption{Cumulative Error Plot}    
            \label{}
        \end{subfigure}
        
        \begin{subfigure}{\columnwidth}   
            \centering 
            \includegraphics[width=0.9\columnwidth,trim={0cm 1cm 0 0},clip]{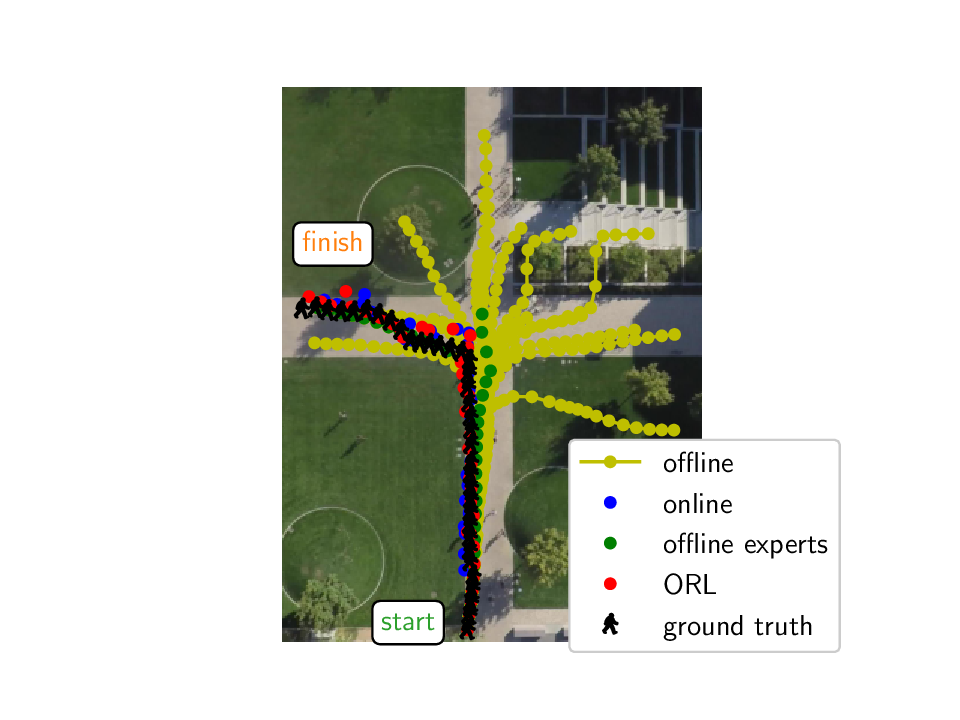}
            \caption{Predicted Trajectories}
        \end{subfigure}
        \hfill
        \caption{Simulation results for the Hyang0 scene. Plots depicting the instantaneous and cumulative errors, showcasing how each prediction method performs over the horizon $T$, are depicted in (a), (b), respectively. In (c), the predicted trajectories, as well as the ground truth one are plotted on a sample photo of the scene. For visualization reasons, trajectories are downsampled by a factor of 30.}
        \label{fig:hyang0}
\end{figure}

\begin{figure}[H]
        \centering
        \begin{subfigure}{0.85\columnwidth}
            \centering
            \includegraphics[width=\columnwidth]{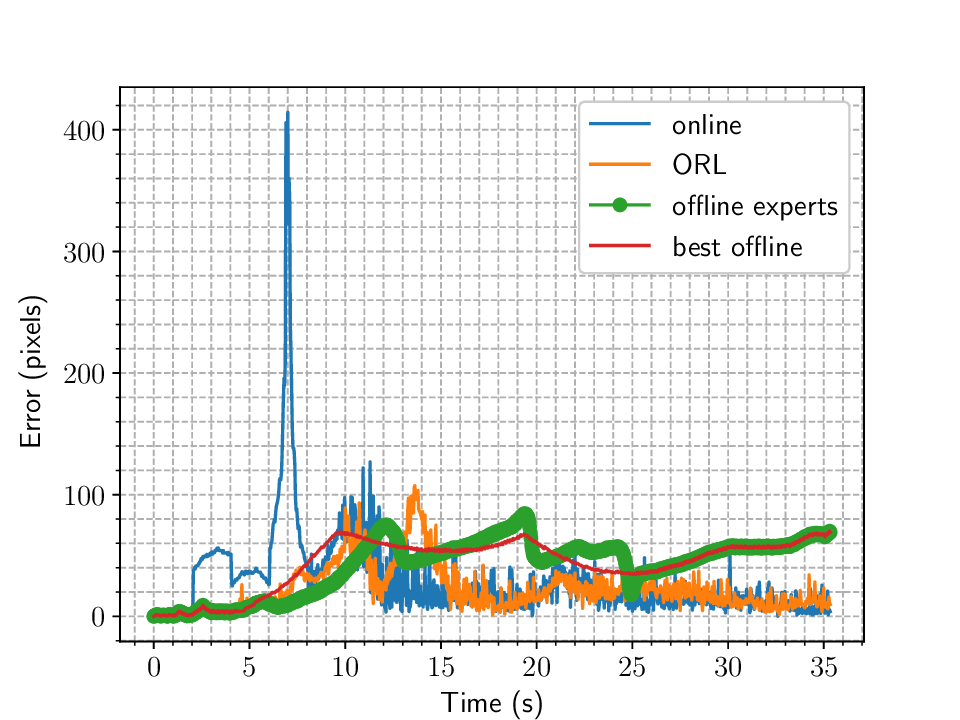}
            \caption{Instantaneous Error Plot}   
            \label{}
        \end{subfigure}

        \begin{subfigure}{0.85\columnwidth}  
            \centering 
            \includegraphics[width=\columnwidth]{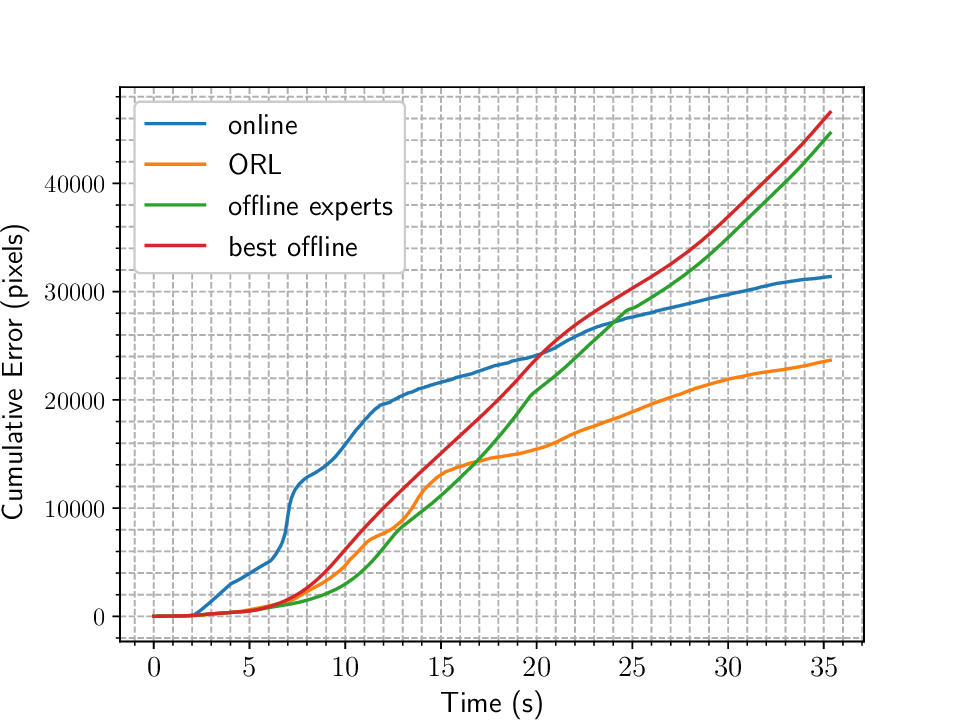}
            \caption{Cumulative Error Plot}    
            \label{}
        \end{subfigure}

        \begin{subfigure}{0.9\columnwidth}   
            \centering 
            \includegraphics[width=\columnwidth,trim={0cm 3cm 0 0},clip]{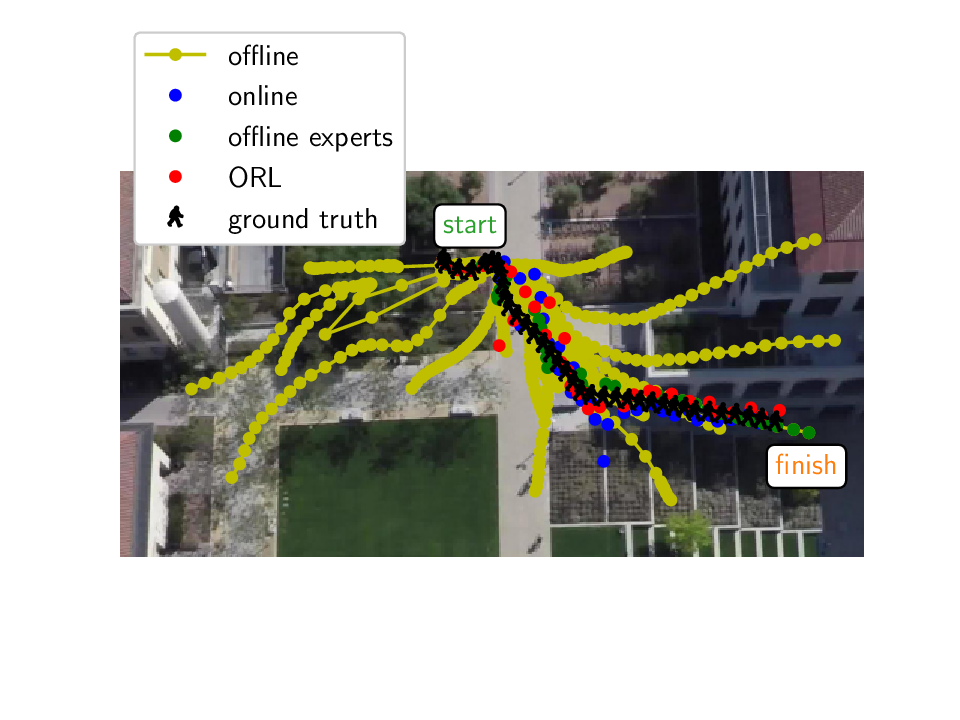}
            \caption{Predicted Trajectories}
        \end{subfigure}
        \hfill
        \caption{Simulation results for the Hyang3 scene. Plots depicting the instantaneous and cumulative errors, showcasing how each prediction method performs over the horizon $T$, are depicted in (a), (b), respectively. In (c), the predicted trajectories, as well as the ground truth one are plotted on a sample photo of the scene. For visualization reasons, trajectories are downsampled by a factor of 30.}
        \label{fig:hyang3}
\end{figure}

The simulation results for the `Hyang0' and the `Hyang3' scenes are shown in \cref{fig:hyang0} and \cref{fig:hyang3}, respectively. In both cases, we consider $N=20$ offline experts and a $k=60$-steps ahead prediction, which corresponds to a $2s$ ahead prediction in the time domain.

The benefits of our method are apparent in the figures; it uses the offline predictions as a ``warm start" to the online prediction and it achieves an improved transient performance. Overall, our method outperforms the purely online method, since the latter lacks the prior expert knowledge that ours possesses. Moreover, it also outperforms the offline experts method, due to the online adaptation that it performs. The results suggest that our method indeed combines the benefits of both offline and online approaches.

\begin{table}[!t]
\caption{Prediction Methods}
\label{tab:pred_methods}
\centering
\begin{tabular}{p{0.45\columnwidth}p{0.45\columnwidth}}
\toprule
\textbf{Prediction Method} & \textbf{Description} \\
\midrule
ORL & Our proposed Online Residual Learning from Offline Experts prediction method.\\
\midrule
online & The purely online prediction method, where we employ the RLS algorithm to predict directly the target state $r_t$, disregarding any prior expert knowledge. \\
\midrule
offline experts & The prediction method where we disregard the residual errors and directly plug the offline predictions in~\eqref{eq:augmented_expert_prediction}. As previously discussed, this case has zero path length. \\
\midrule
best offline & The best offline prediction that the model of \cite{Mangalam_2021_ICCV} produces, with respect to minimizing the ADE. \\
\bottomrule
\end{tabular}
\end{table}

%% file: sections/appendix.tex
\subsection{Proof of~\cref{thm:regret_guarantees}}\label{proof:thm:regret_guarantees}

First, we provide the definition of \emph{exp-concavity} and a result regarding~\cref{alg:meta_alg}, adopted from \cite{cesa2006prediction,yuan2020trading}

\begin{definition}
    A function $f:\mathcal{D}\to\mathbb{R}$ is \emph{$\alpha$-exp-concave} if the function $\exp(-\alpha f(x))$ is concave for some $\alpha > 0$.
\end{definition}

It turns out that if the framework of multiple experts is applied to exp-concave losses, then the regret with respect to the best experts scales with $\log N$, where $N$ is the number of experts. The following result is standard see, for example,~\cite{cesa2006prediction,yuan2020trading}.
\begin{lemma}[Multiple experts, exp-concavity]\label{lem:MOE}
Define the cumulative losses $\mathcal{L}_{T,i} \triangleq \sum_{t=1}^T \ell_t(\hat{r}_{t,i})$ for every expert $i\in\{1,\dots,N\}$.    Let the loss function $\ell_t$ be $\alpha$-exp-concave, with $\alpha > 0$. Set the learning rate $\lambda=\alpha$. Then, the cumulative loss of~\cref{alg:meta_alg} is bounded by
    \[
    \mathcal{L}_T \leq \min_{i=1,2,\dots,N}\mathcal{L}_{T,i} + \frac{1}{\alpha} \log N .
    \]
\label{thm:multi_experts}
\end{lemma}

If the number of experts is independent of the time horizon, then, this logarithmic term is negligible on average.

In our case, we have squared norm losses, which are exp-concave under boundedness conditions. The following property holds, which is standard. See for example Lemma 2.3 in~\cite{foster2020logarithmic}
\begin{lemma}[Exp-Concavity of square loss]\label{lem:exp_concavity}
    Consider the function $f(x) = \norm{x-y}^2 $. Then, $f$ restricted to $\norm{x-y}^2\le C$, for some $C>0$ is $(2C)^{-1}$-exp-concave. 
\end{lemma}

In our case, from~\cref{assum:bounded_residuals}, we have that $e_{t,i}$ is bounded, i.e. $\norm{e_{t,i}}\leq D_r$, for all $i\in\{1,2,\dots,N\}$, $1\leq t \leq T$. Moreover, the predictions $\hat{e}_{t,i}$ are also bounded due to the choice of constraint set $S$, $\norm{\hat{e}_{t,i}} \leq D D_r$ for all $i\in\{1,2,\dots,N\}$, $1\leq t \leq T$. In particular, we always have
$\norm{e_{t,i}-\hat{e}_{t,i}}^2\le 2D^2_r+2D^2D^2_r$. Hence, by Lemma~\ref{lem:exp_concavity}, the square loss $\norm{r_t - \hat{r}_{t,i}}^2=\norm{e_{t,i} - \hat{e}_{t,i}}^2$ is $\alpha$ exp-concave, for $\alpha=(4D^2_r+4D^2D^2_r)^{-1}$.

We can now finish the proof of~\cref{thm:regret_guarantees}:

\begin{proof}
Invoking ~\cref{lem:MOE} for $\lambda=(4D^2_r+4D^2D^2_r)^{-1}$, we obtain
\[
 \mathcal{L}_T \leq \min_{i=1,2,\dots,N}\mathcal{L}_{T,i} + \frac{1}{\alpha} \log N.
\]
Now fix an expert $i$.
    By Theorem~4.5 of~\cite{tsiamis2024predictive} (and following the same procedure as Corollary~4.2 therein) if $\gamma_i$ is chosen as in \eqref{eq:forgetting_factor}, then 
    \begin{multline}\label{eq:rls_dr}
\mathcal{L}_{T,i} -\mathop{\min_{M_{t,i}\in\mathcal{S}}}_{V_{T,i}\le V_T}\mathcal{L}_T(M_{1:T,i})    \\  \leq \max\{\mathcal{O}(\log T), \mathcal{O}(\sqrt{TV_{T}}).\}
    \end{multline}
    Putting everything together completes the proof.
\end{proof}

\subsection{Multiple-steps ahead prediction} \label{app:k-steps ahead RLS}
\begin{algorithm}
	\caption{$k$-steps ahead RLS prediction}
	\label{alg:k_step_RLS}
	\begin{algorithmic}[1]
		\Require Forgetting factor $\gamma_i \in (0,1]$ for expert $i\in\{1,2,\dots,N\}$, regularizer $\varepsilon_i$
		\State Initialize k learners $\hat{M}_{j+k, i}\in S, j=0,1,\dots,k-1$ 
		\State Initialize $P_{j,i} =\varepsilon_i I_{np}, j=0,1,\dots,k-1$
		\For{$t=k,\dots,T$} 
		\State Make prediction $\hat{e}_{t,i} = \hat{M}_{t,i}z_{t-k,i}$
		\State Receive true state $r_t$, compute residual error $e_{t,i}$ 
		\State Incur loss $\ell_t(\hat{M}_{t,i}) = \norm{e_{t,i} - \hat{e}_{t,i}}^2$
		\State Update $P_{t,i} = \gamma_i P_{t-k, i} + z_{t-k,i} z_{t-k,i}^{\top}$
		\State $\hat{M}_{t+k, i}^{\star} = \hat{M}_{t, i} + (e_{t,i} - \hat{M}_{t, i} z_{t-k, i})z_{t-k, i}^{\top} P_{t, i} ^{-1}$
		\State Project $\hat{M}_{t+k, i} = \arg \min_{M\in S}\norm{M-\hat{M}_{t+k, i}^{\star}}_{F, P_{t,i}}$
		\EndFor
		
	\end{algorithmic}
\end{algorithm}
Our approach applies to the general case of $k$-steps ahead prediction as well. Let $\hat{r}_{t+k}$ represent the prediction of $r_{t+k}$ at time $t$.
The main difference is that we face \emph{delayed feedback}~\cite{joulani2013online}, that is, the true target state $r_{t+k}$ is revealed after $k$ steps into the future. To derive a result similar to Theorem~\ref{thm:regret_guarantees} we need to slightly modify Algorithm~\ref{alg:RLS} while we keep Algorithm~\ref{alg:meta_alg} the same. In particular, the guarantees for the RLS algorithm~\cite{yuan2020trading} do not apply directly. However, using the black box reduction of~\cite{joulani2013online}, we can turn the delayed feedback problem into a non-delayed one~\cite{tsiamis2024predictive}. 

For each expert $i$, $i=1,\dots,N$, we keep $k$ independent copies of the linear model $\hat{M}_{t+k,i}$, which are updated at non-overlapping time steps.

At each time step $t\ge 0$, only the $j-$th learner, for $j=t\bmod k\in\{0,\dots,k-1\}$, is activated.  We minimize the cost:
\begin{equation}
\begin{aligned}
 \hat{M}_{t+k,i}&=\arg\min_{M\in\mathcal{S}}   J_{t,k}(M)+\gamma_i^{\lfloor t/k\rfloor}\varepsilon_{i} \norm{M}_F^2,\\
 J_{t,k}(M)&\triangleq \sum_{\tau=0}^{\lfloor t/k\rfloor-1} \gamma_i^{\tau} \norm{e_{t-\tau k,i} - M z_{t-(\tau+1)k,i}}^2,
    \label{eq:cost_k_steps_ahead}
    \end{aligned}
\end{equation}
where $\varepsilon_i, \gamma_i, S$ are the same as in \eqref{eq:cost}. Observe that $\hat{M}_{t+1,i},\dots,\hat{M}_{t+k,i}$ are updated based on non-overlapping data sets. Similarly to the $1$-step ahead prediction, each expert predicts
\begin{equation}\label{eq:k_step_RLS_prediciton}
\hat{e}_{t+k,i}=\hat{M}_{t+k,i}z_{t,i}.
\end{equation}
The minimization problem \eqref{eq:cost_k_steps_ahead} enjoys a recursive solution with projections, which is summarized in~\cref{alg:k_step_RLS}

Following the steps of~\cite{tsiamis2024predictive}, we can obtain guarantees similar to Theorem~\ref{thm:regret_guarantees}. The regret will be of the order of
\[
\max\{\mathcal{O}(\log T), \mathcal{O}(\sqrt{TV_{T}})\}
\]
where now $V_T$ is the upper bound on the $k$-step ahead path lengths
\[
V_{T,i}^k = \sum_{t=k}^{T-k} \norm{M_{t+k,i} - M_{t,i}}_F.
\]